\documentclass[sigconf,  screen]{acmart}

\AtBeginDocument{%
  }
    
\usepackage{algorithmic}
\usepackage{graphicx}
\usepackage{textcomp}
\usepackage{tabularx}
\usepackage{enumitem}
\usepackage{array}
\usepackage{booktabs}
\usepackage[table]{xcolor}  
\usepackage{tcolorbox}      
\usepackage{url}
\usepackage{balance}

\usepackage{hyperref}       

\begin{document}

\copyrightyear{2026}
\acmYear{2026}
\setcopyright{cc}
\setcctype{by}
\acmConference[ICSE-NIER '26]{2026 IEEE/ACM 48th International Conference on Software Engineering}{April 12--18, 2026}{Rio de Janeiro, Brazil}
\acmBooktitle{2026 IEEE/ACM 48th International Conference on Software Engineering (ICSE-NIER '26), April 12--18, 2026, Rio de Janeiro, Brazil}
\acmPrice{}
\acmDOI{10.1145/3786582.3786828}
\acmISBN{979-8-4007-2425-1/2026/04}

\title{Towards Bridging Language Gaps in OSS with LLM-Driven Documentation Translation}
\author{Elijah Kayode Adejumo}
\affiliation{%
  \institution{George Mason University}
  \city{Fairfax}
  \country{USA}}
\email{eadejumo@gmu.edu}
\author{Mariam Guizani}
\affiliation{%
  \institution{Queen's University}
  \city{Ontario}
  \country{Canada}}
\email{mariam.guizani@queensu.ca}
\author{Fatemeh Vares}
\affiliation{%
  \institution{George Mason University}
  \city{Fairfax}
  \country{USA}}
\email{fvares@gmu.edu}
\author{Brittany Johnson}
\affiliation{%
  \institution{George Mason University}
  \city{Fairfax}
  \country{USA}}
\email{johnsonb@gmu.edu}

\begin{abstract}
While open source communities attract diverse contributors across the globe, only a few open source software repositories provide essential documentation, such as \textsc{ReadMe} or \texttt{CONTRIBUTING} files, in languages other than English. 
Recently, large language models (LLMs) have demonstrated remarkable capabilities in a variety of software engineering tasks.
We have also seen advances in the use of LLMs for translations in other domains and contexts.
Despite this progress, little is known regarding the capabilities of LLMs in translating open-source technical documentation, which is often a mixture of natural language, code, URLs, and markdown formatting. 
To better understand the need and potential for LLMs to support translation of technical documentation in open source, we conducted an empirical evaluation of translation activity and translation capabilities of two powerful large language models (OpenAI's ChatGPT 4 and Anthropic's Claude). 
We found that translation activity is often community-driven and most frequent in larger repositories.
A comparison of LLM performance as translators and evaluators of technical documentation suggests LLMs can provide accurate semantic translations but may struggle preserving structure and technical content.
These findings highlight both the promise and the challenges of LLM-assisted documentation internationalization and provide a foundation towards automated LLM-driven support for creating and maintaining open source documentation.

\end{abstract}



\maketitle
\section{Introduction \& Background}
Users, contributors, and maintainers all play an integral part in the success and sustainability of open source innovation, which means providing valuable support for all roles is beneficial. Written documentation serves as a fundamental source of support in open source projects \cite{carvalho2013open}. One vital piece of documentation in open source repositories is the \textsc{ReadMe} file usually in markdown format \cite{gruber2012markdown}, which plays a vital role in introducing new users and potential contributors to the software project, making its content essential \cite{prana2019categorizing}. The importance of the \textsc{ReadMe} file has attracted attention from researchers who evaluate how these files are created \cite{liu2022readme}, how useful they are to users and contributors, and which techniques maintainers can adopt to improve them.
Recent findings \cite{venigalla2024there,wang2023study,prana2019categorizing} have shown a positive correlation between a repository's \textsc{ReadMe} file content and various repository meta-characteristics such as popularity, development progress, and community size, emphasizing the importance of high-quality, accessible \textsc{ReadMe} files across repositories. Another study \cite{SummarizationOfReadme} proposed the use of transformers to summarize extremely long \textsc{ReadMe} files, thereby supporting better comprehension. While the ability to leverage the \textsc{ReadMe}, and other documentation, is important and valuable to all parties, studies suggest existing documentation may not be accessible to diverse audiences~\cite{carvalho2013open, adejumo2024towards}. Prior work 
revealed challenges faced by contributors due to a lack of documentation in preferred languages and that contributors who are not confident in English are more likely to report struggles in following discussions \cite{guizani2021long,guizani2022perceptions}. With the growing contribution index from non-English speaking countries all over the globe ~\cite{GitHubReport} there is an even more pressing need to understand and improve the current state of documentation accessibility.


In recent years, large language models (LLMs) have demonstrated significant strengths in supporting various aspects of software engineering~\cite{zheng2025towards, hou2024large, 10449667}. 
A recent mining study revealed the use of ChatGPT to improve \textsc{ReadMe} files and other relevant documentation~\cite{tufano2024unveiling}.
Prior work has also indicated the potential for LLMs to improve, simplify, and summarize open source documentation to reduce information overload and increase the readability for contributors~\cite{adejumo2024towards, 10.1145/3593856.3595910, tufano2024unveiling}.
Moreover, LLM-driven translation and quality estimation have demonstrated state-of-art performance outside of computing~\cite{freitag2021results, freitag2022results, kocmi-federmann-2023-large, huang2023towards, he2024exploring, moslem-etal-2023-adaptive, calzi2025science}; in fact, prior work suggests that with proper prompting strategies LLMs can achieve translation quality comparable to that of human experts \cite{zhang2023prompting}. Given LLMs' joint training on natural-language and code corpora, we hypothesize that modern LLMs are better suited than conventional machine translation pipelines to handle the mixed Markdown + code + URL structure found in technical documentation. Although all of these insights are promising, there is little evidence to support the ability of LLMs to be used to translate technical documents such as \textsc{ReadMe} files.

To this end, we investigated \textsc{ReadMe} translation contributions across repositories of different scales. 
By mining all pull requests related to \textsc{ReadMe} translations, we found signals that indicate translation activity, though often community-driven, and a need for translation support.
To evaluate the capabilities of LLMs for providing technical documentation translation support, we sampled 50 \textsc{ReadMe} files from large repositories ($>10000$ stars and forks) and evaluated LLM-generated translations with ChatGPT and Claude.
Our LLM-based evaluation suggested high semantic accuracy, however, a manual evaluation revealed significant preservation challenges: both LLMs struggled with maintaining URLs, markdown formatting, and code snippets. 
These insights have implications for pipelines that leverage LLMs for translation of technical artifacts, such as the need for mechanisms to assess and improve structural preservation in this context.
Building on this foundation, our future efforts will investigate 1) the development of a framework for measuring structural preservation in technical documentation translations, 2) fine-tuning of LLMs with structure-aware objectives that penalize corruption of code, links, and markdown, and 3) perspectives from experts and open source contributions regarding the ability for our approach to provide practical support.


 
Our initial focus is on identifying translation patterns across open-source software repositories and evaluating LLMs' translation capabilities for technical documentation. With this, our efforts described in this paper aim to answer the following research questions:

\begin{description}
    \item[\textbf{RQ1}]\textit{What patterns of translation contribution in open-source repositories reveal the global significance for multilingual technical documentation, and how effective and sustainable are these contributions?}
    \item[\textbf{RQ2}]\textit{How effective are LLMs technologies ChatGPT and Claude in translating technical documentation such as README files with respect to technical accuracy, preservation of code/markdown formatting?}
\end{description}

\section{Methodology}\label{sec:dataset}
To reveal translation activities across repositories, we analyzed repositories at three scales based on stars and forks, as these metrics typically represent community engagement and project adoption \cite{borges2018s,jiang2017and}.

We divided repositories into \textbf{Small-Scale:} 100 - 5,000 stars and forks, \textbf{Medium-Scale:} 5,000 - 10,000 stars and forks, and \textbf{Large-Scale:} Stars and forks greater than 10,000.

We focused on the README file due to its role as the primary introductory document that users encounter when accessing repositories \cite{gaughan2025introduction}, we considered it a strong foundation for understanding translation issues  and contribution patterns. For each repository category (small-scale, medium-scale, and large-scale), we searched for and extracted pull requests and issues whose titles contained ``translate README'' using a keyword-based approach \cite{yu2022keyword}. This allowed us to assess the current state of translation contributions and issues raised across different repository categories. We then analyzed the timestamps of these pull requests to understand contribution trends and identify when translation activity peaks, providing insights into the periods or seasons during which translation demand is most likely to surge. To evaluate how effective and sustainable these translation contributions are, we examined whether the pull requests were actually merged and, if merged, how well they kept up with their rapidly evolving English source. We compared the rate of change in the English and translated versions 180 days (six months) after each pull request was merged.




\subsection{LLM Translation Process}

To explore the feasibility of LLM-powered technical documentation 
translation, we selected  \textsc{ReadMe} files from the top 50 large-scale repositories, as these projects are widely adopted by developers worldwide \cite{jiang2017and}. \textbf{Translation Process:}
As a first step, we focused on translation from English to German based on a recent report from GitHub \cite{GitHubBlog} that indicated Germany ranks among the top three contributors by region outside of the US.
In addition, various LLM translation and evaluation studies  have also evaluated this language translation,  indicating that German is a high-resource language \cite{volk2024llm, kocmi2024findings, koshkin2024transllama}. 

We provided the extracted markdown files to the LLMs (ChatGPT-4o and Claude Sonnet) using the following prompt: \\

\textit{ \textbf{``You are an expert translator, please translate the following Markdown content from English to German. 
Preserve code blocks, links and markdown format exactly as they are.''}}\\


In recent years, LLM prompting has been evolving, with studies \cite{chen-etal-2023-mapo} showing that prompt optimization techniques such as chain-of-thought reasoning, zero, few, and one-shot prompting can  improve results generated by LLMs. As we move towards our vision for automated technical documentation translation powered by LLMs, other prompting strategies would be employed.

\begin{figure*}[ht]   
    \centering
    \includegraphics[width=5in]{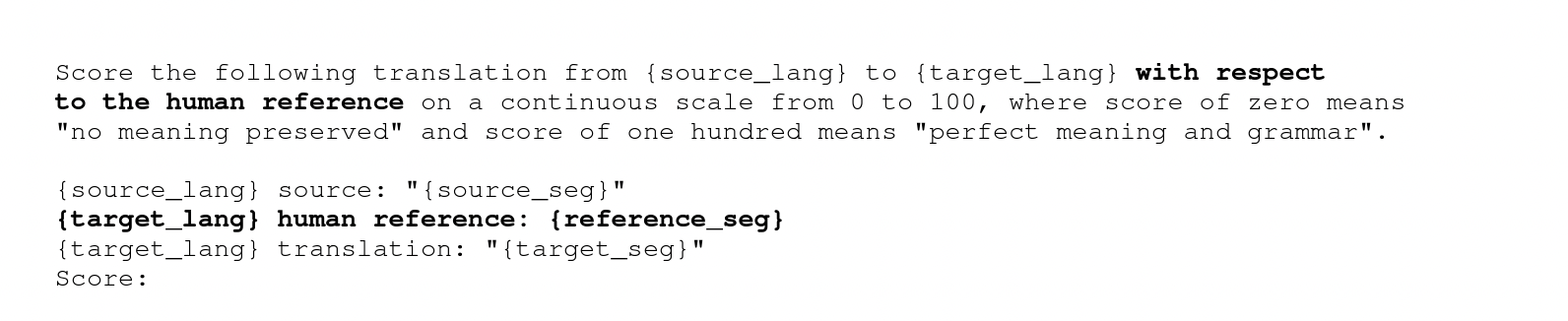}
    \caption{Translation scoring prompt template}
    \label{fig:prompt-template}
\end{figure*}

\subsection{Translation Evaluation}

Recent studies have shown that LLMs can achieve state-of-the-art machine translation \cite{hendy2023good} for high-resource languages. A more recent study \cite{kocmi-federmann-2023-large} argued that if LLMs are capable of performing state-of-the-art machine translation, they should also be able to provide state-of-the-art evaluation of translation quality. To support this claim, the authors introduced GEMBA (GPT Estimation Metric Based Assessment), a GPT-based metric for assessing translation quality.
Building on these efforts, we evaluated our translation quality by conducting a cross-model translation evaluation, which involves using one LLM to generate the translation and another LLM to evaluate the translation (in our case, ChatGPT 4.0 and Claude 3.5 Sonnet, respectively), while employing the GEMBA assessment framework.
We utilized GEMBA-Direct Assessment where LLMs are provided a reference and GEMBA-Direct Assessment [noref] where the LLMs are not provided a reference to guide their scoring \cite{kocmi-federmann-2023-large}. As shown in Figure~\ref{fig:prompt-template}, the human reference in this context is the original \textsc{ReadMe} file.

\subsubsection{Manual Validation}
 To evaluate structural preservation of code, URLs, and markdown formatting, we (the authors) manually evaluated the structure of the translated files in comparison to their English counterparts to determine whether these key fidelity components were preserved after translation.

\section{Preliminary Findings}
\subsection{\textbf{Patterns of Translation (RQ1) }}

Our findings indicate  translation activities across the open-source software community. These translations spanned across multiple languages, demonstrating that global contributions are increasingly significant, as supported by GitHub's recent report \cite{GitHubReport}. Our replication package \cite{replication_package} contains the extraction file showing the distribution of languages observed. We also observed some attempts to improve translation quality. In one notable case, an Italian speaker reported an issue indicating that the existing translation felt mechanical and lacked natural flow \cite{GitHubTranslationReport}.

\textbf{Pull Requests:}
After mining pull requests with ``Translate ReadMe'' included in the title, we found that for \textbf{large-scale repositories} (Section~\ref{sec:dataset}), there were 428 pull requests across 254 repositories, and some repositories had more than one translation request. 
These pull requests included translations into multiple languages and updates to existing translations, reflecting a community-driven translation effort. 
For \textbf{medium-scale repositories}, we found 13 pull requests across 58 repositories, and for \textbf{small-scale repositories}, we found 24 pull requests across 500 repositories. 
Another relevant trend we observed with the timestamps of these pull request  from 2014 to 2025 was that about 30\% of all translation pull requests occurred in the month of October across large-scale repositories, which we believe may have been influenced by Hacktoberfest ~\cite{hacktober}, a month-long global open source contribution event. Figure \cite{Translation_activity_trend} shows the patterns in translation activity trends throughout each year from 2014-2025.
\textbf{Issues:}
We also examined issues in repositories that included ``Translate ReadMe'' in their titles to see how frequently users and contributors raised related concerns. 
A similar trend emerged: In \textbf{large-scale repositories}, we found 64 issues across 260 repositories, in \textbf{medium-scale repositories}, 3 issues across 57 repositories; and in \textbf{small-scale repositories}, 4 issues across 500 repositories. The distribution of these issues is shown in Figure S1 of our replication package
\cite{Translation_Issues_activity_trend} 

These results reveal disparities in translation activities across the life-cycle of open source repositories. Although translation efforts occur across levels of maturity, we observed a clear concentration of translation activities in large-scale repositories, though primarily opportunistic and likely influenced by project maturity and/or visibility. 
Table \ref{tab:translation_stats} shows the distribution of translation activities.

\begin{table}[ht]
\centering
\caption{Translation Activity Statistics by Repository Scale}

\label{tab:translation_stats}
\begin{tabular}{lccc}
\toprule
\textbf{Repository Scale} &  \textbf{PRs} & \textbf{PRs merged} & \textbf{Issues} \\
\midrule
Large Scale ($>$ 10k)  & 428 & 61.0\%  & 64 \\
Medium Scale (5k-10k)  & 13 & 84.46\% & 3 \\
Early Stage (100-5k) & 24  &70.8\% & 4 \\
\bottomrule
\end{tabular}
\end{table} 

\subsubsection{Effectiveness \& Sustainability of Translation PRs}
Across the  \textbf{Large-scale} repositories we found 428 translation pull-requests.  
Most were accepted: 261 merged (61\,\%) vs. 167 closed.  
Table~\ref{tab:adoption_rate_large} summarises per-repository adoption.


To understand post-merge upkeep in a six-month activity window, we sampled  merged  pull requests and evaluated if they are able to keep up with their English counterpart, that is evaluating if a merged pull request language-tagged variant such as \texttt{README.fr.md}, \texttt{README-zh\_CN.md} are able to keep up with changes to  \texttt{README*.md}. We found that the English sources kept evolving (median 8.5 commits) whereas the corresponding translations received \emph{no} follow-up commits, yielding a maintenance gap. Table~\ref{tab:survival_large_repo} shows the distribution.


\begin{table}[ht]
\centering
\caption{Adoption Ratio (Merged PRs/Total PRs) among 69 large repositories. 
}
\label{tab:adoption_rate_large}
\begin{tabular}{lc}
\toprule
\textbf{Metric} & \textbf{Value} \\
\midrule
Mean repo–level adoption rate & 0.43 \\
Median repo–level adoption rate & 0.50 \\
\bottomrule
\end{tabular}
\end{table}



\begin{table}[ht]
\centering
\caption{Six-month activity after merged translation PRs in Large Repositories (20 cases)}
\label{tab:survival_large_repo}
\begin{tabular}{lcc}
\toprule
\textbf{Metric} & \textbf{English} & \textbf{Translation} \\
\midrule
Median commits & 8.5 & 0 \\
Mean commits & 23.6 & 4.2 \\
\midrule
\multicolumn{3}{c}{\textbf{Commit Gap}} \\
\midrule
Maximum English commits ahead & \multicolumn{2}{c}{166} \\
\bottomrule
\end{tabular}
\end{table}

\smallskip
For \noindent\textbf{medium-scale} repositories we observed 13 PRs (11 merged); no repository showed a maintenance gap, that is, the English READMEs remained unchanged during the six-month activity window.

In \textbf{small-scale} repositories, we identified 24 translation PRs, of which 17 were merged. Six translations lagged behind their English counterparts.

\subsection{\textbf{Translation Quality Evaluation (RQ2) }}

LLM-based Translation Evaluation: utilizing  GEMBA scoring framework \cite{kocmi-federmann-2023-large}, our findings indicate that both models provide reasonable assessments of translations. Specifically, Claude scored GPT's translations at an average of 94–95\% (with and without referencing the original README file), while ChatGPT scored Claude's translations at an average of 94–96\% across the 50 assessed \textsc{ReadMe} files. We observed that ChatGPT scored Claude's translations higher when we provided reference \textsc{ReadMe} files, whereas there was no difference in Claude's evaluation of ChatGPT's translations (with or without reference). 
Our cross-model evaluation procedure demonstrates that LLMs can provide reasonable quality assessments of translated technical documentation even without referencing the original documentation. 

Manual Validation: Although automated evaluation indicated high semantic quality with scores above 94\%, our manual evaluation of all 50  \textsc{ReadMe}  files revealed inconsistencies in the preservation of code, URLs, and markdown formatting. We found that 
Claude produced inconsistencies in 16\% of files (8/50), while ChatGPT  translations showed inconsistencies of about 50\% (25/50). URL inconsistencies between the original and translated versions were the most prevalent, occurring in 14\% of Claude translations (7/50) versus 32\% for ChatGPT (16/50). Code formatting issues such as missing code snippets or code fractions after translation appeared in 2\% of Claude outputs (1/50) compared to 18\% for ChatGPT (9/50), while markdown formatting problems occurred in 8\% (4/50) and 24\% (12/50) respectively. Some files exhibited multiple inconsistency types simultaneously. This manual validation reveals that automated scoring may not adequately capture structural fidelity issues, suggesting that high semantic quality scores can coexist with significant structural preservation problems in technical documentation translation.


\section{Discussion \& Future Work}
We found that community engagement often plays a key role in supporting documentation translation within open source projects.
However, our findings indicate that this support primarily benefits larger, more established repositories as 
smaller repositories often lack the visibility or contributor base needed for consistent translation efforts~\cite{miller2019people}.
Even in more mature projects that do receive occasional contributions, sustaining and updating translations remains a significant challenge.
LLMs present a promising, yet underexplored, alternative by offering on-demand translation capabilities that are independent of community size, event timing, or project popularity.

While our findings suggest LLM-generated translations may not yet be fully reliable on their own, they can serve as a strong foundation for a more systematic, AI-assisted approach to multilingual documentation.\
In cases where no translation currently exists, LLMs can produce initial drafts, easing the effort required to introduce diverse language support.
For projects that update their original (typically English) documentation, change-detection tools could prompt LLMs to generate updated translations automatically.
Moreover, as our findings indicate, providing LLMs with access to prior translations can enhance both quality and consistency, offering a path to improve translation updates over time.



\textbf{Improving LLM Translation Quality.} Our findings indicate that LLMs are capable of providing quality semantic translations of documentation. However, deeper inspection revealed that LLMs may struggle with preserving structural aspects of technical documentation such as the formatting, in-line code snippets, and urls.
Unlike general-purpose translation tasks, technical documentation requires attention to structural fidelity, accurate technical terminology, and completeness.
As interest in LLMs grows, research increasingly supports the need for fine-tuning in domain-specific contexts~\cite{huang2023empirical, tinn2023fine, bakker2022fine}.
A promising direction for improving translation of technical documentation lies in adapting pre-trained LLMs such as ChatGPT and Claude specifically for this purpose.
This could involve designing loss functions that penalize structural deviations during training and incorporating verification mechanisms to ensure all sections from the source documentation are preserved in the output.





\textbf{Alignment \& Localization of Translations.}
Despite fidelity issues, our findings suggest LLMs are capable of providing accurate semantic translations. 
However, prior work suggests it is important to consider localization-the process of recognizing differences in needs, dialects, and culture of an audience-when providing translations of technical documentation to ensure quality~\cite{ledet2005following, adejumo2024towards}.
Even a carefully tuned framework or LLMs may not capture every nuance that makes a translation usable for developers.
To ensure alignment of translated technical material, it is important to engage regional expertise to better understand the needs and considerations in generating localized content~\cite{ledet2005following}.
Therefore, one critical direction for realizing the benefits of documentation translations is to better understand the process of localization and patterns that can facilitate consideration in practice.

\textbf{Future Work.}
Our efforts thus far have provided valuable and actionable insights on which we plan to build.
First, while fine-tuning the LLM or the prompt can improve outcomes, the burden still lies on the user (in this case, contributors or maintainers) to ensure preservation of content and structure.
Therefore, we are working to develop a preservation scoring framework that can support the assessment of documentation translation for fidelity issues. 
We will begin by developing the framework to support detection of preservation issues with urls, markdown formatting, and code snippets. 
We will evaluate our initial framework on the dataset used in this paper to ensure it accurately measures preservation.
To ensure the robustness of our framework, we will extend our automated and manual evaluation to more repositories and extend our framework as needed. Another important consideration is the ability for our framework to integrate into into existing open source pipelines and workflows.

Using all these insights, we will extend our pipeline to support automated updates of translated documentation as original (English) documentation changes.

\balance
\bibliographystyle{ACM-Reference-Format}
\bibliography{references}
 
\end{document}